\RequirePackage{lineno}
\documentclass[linenumbers,author-year,preprint,showkeys,preprintnumbers,amsmath,amssymb,superscriptaddress]{revtex4}
\usepackage{graphicx}
\usepackage{graphics}
\usepackage{subfigure}
\usepackage{dcolumn}
\usepackage{bm}
\usepackage{txfonts}
\usepackage{enumerate}
\usepackage{amsmath,empheq,mathrsfs}

\makeatletter
\newcommand*{\rom}[1]{\expandafter\@slowromancap\romannumeral #1@}
\makeatother
\begin{document}


\centerline{Social learning of prescribing behavior can promote population optimum of antibiotic use}	

{\small {\vskip 12pt \centerline{Xingru Chen$^{1}$, Feng Fu$^{1,2}$}

\begin{center}
$^1$ Department of Mathematics, Dartmouth College, Hanover, NH 03755, USA\\
$^2$ Department of Biomedical Data Science, Geisel School of Medicine at Dartmouth, Lebanon, NH 03756, USA
\end{center}
}}

\vskip 30pt

\begin{minipage}{142mm}
\begin{flushleft}

{\textbf{Running title:}}\, Antibiotic Resistance Game
{\textbf{Manuscript information:}}\, Original article.\\
{\textbf{Author contributions:}}\, X.C., \& F.F. conceived the model, performed analyses, and wrote the manuscript.\\
{\textbf{Corresponding author:}} \\
Feng Fu\\
\small {Department of Mathematics, Dartmouth College\\
27 N. Main Street, 6188 Kemeny Hall\\
Hanover, NH 03755 USA\\
Email: fufeng@gmail.com\\
Tel: +1 (603) 646 2293\\
Fax: +1 (603) 646 1312\\}
\end{flushleft}
\end{minipage}

\clearpage

\begin{center}
{
\begin{minipage}{142mm}
{\bf Abstract:} \, 

The rise and spread of antibiotic resistance causes worsening medical cost and mortality especially for life-threatening bacteria infections, thereby posing a major threat to global health. Prescribing behavior of physicians is one of the important factors impacting the underlying dynamics of resistance evolution. It remains unclear when individual prescribing decisions can lead to the overuse of antibiotics on the population level, and whether population optimum of antibiotic use can be reached through an adaptive social learning process that governs the evolution of prescribing norm. Here we study a behavior-disease interaction model, specifically incorporating a feedback loop between prescription behavior and resistance evolution. We identify the conditions under which antibiotic resistance can evolve as a result of the tragedy of the commons in antibiotic overuse. Furthermore, we show that fast social learning that adjusts prescribing behavior in prompt response to resistance evolution can steer out cyclic oscillations of antibiotic usage quickly towards the stable population optimum of prescribing. Our work demonstrates that  provision of prompt feedback to prescribing behavior with the collective consequences of treatment decisions and costs that are associated with resistance helps curb the overuse of antibiotics.

\vspace*{1\baselineskip}

\end{minipage}
}
\end{center}

\clearpage

\section{Introduction}

Antibiotics have been used primarily as human medicine for the treatment and prevention of bacterial infections for about 80 years; later as a growth promoter applied in animal feeds, for about 65 years~\cite{Jones_PS03, Castanon_PS07}. In this period, it has proved itself incredibly powerful to benefit individual patients, to suppress the overall epidemic of diseases and also to expand livestock production~\cite{Aminov_FIM_10,Zaffiri_JIS12,Boeckel_Science17}. However, the wide use of antibiotics in our society is tagged along by the development of resistance, first identified in the 1940's~\cite{Garrett94,Levy_SA98,Walsh_Nature00,Palumbi_Science01,Levy02,Norrby_LID05,Spellberg_CID08,Arias_NEJM09,Aminov_FIM_10,Davies_MMBR10,Zur_LancetID11,Rossolini_COP14,Cully_Nature14,WBG_book,Reardon_NN17}. 

In recent years, the number of new antibiotics approved by the U.S. Food and Drug Administration (FDA) has been dramatically reduced, suggesting an `EROOM' law (a phenomenon in contrast to the Moore's law)~\cite{Projan_COM03,Norrby_LID05,Scannell_Nature12}. Even worse, the time period of an antibiotic's effectiveness from its introduction to first resistance identified becomes increasingly short (Fig. 1A). Moreover, superbugs (multi-drug resistant bacteria) such as Methicillin-resistant \textit{S. aureus} (MRSA) seem to outsmart our efforts to treat infectious diseases~\cite{Garrett94,Flores_CCJM96,Levy_SA98,Walsh_Nature00,Spellberg_CID08,Arias_NEJM09,Woodford_JI09,Conly_BWHO10,Magiorakos_CMI12,Rossolini_COP14}. Antibiotic resistance is associated with worsening mortality and medical costs~\cite{CDC13}. As a consequence, we are confronted with antibiotic resistance crisis, at the risk of running out of effective antibiotics for infection treatments~\cite{Reardon_NN17}. 

One of the important factors contributing to the the fast emergence of resistance is overprescribing~\cite{Nyquist_Jama98,Christakis_MS11,Hallsworth_Lancet16,Murshid_PP17}. High demand for antibiotics driven by individual self-interest is not necessarily aligned with the social optimum of antibiotic consumption. Under certain conditions, the overuse of antibiotics can lead to the tragedy of the commons~\cite{Hardin_JNRPR09,Baquero_REQ03,Foster_PLoS06,Conly_BWHO10,Porco_PLoS12}. Therefore, it is of significant public health interest to understand and manage antibiotic resistance from this behavioral perspective. 

Here, we focus on the interaction of prescription behavior and resistance evolution through a feedback loop (Fig. 1B): collective outcomes of prescribing decisions affect the underlying resistance evolution, which in turn influences prescription behavior. The behavior-disease interaction model of this kind is simple yet proof of concept, and sheds light on how social learning of prescription behavior in response to the underlying evolutionary dynamics of resistance can render population optimum of antibiotic use.

\section{Results \& Discussion}

To begin with, we use an evolutionary epidemiological model to describe the competition dynamics of sensitive versus resistant strains (see Methods \& Model). We focus on quantifying the extent to which the (over)use of antibiotics would cause the emergence of resistance in the long run. To do so, we introduce the parameter $\theta$ to denote the presentation rate of infected individuals who bring their condition to a physician's attention and seek antibiotic treatment for their illness (In this regard, the $\theta$ value is determined by individual disease awareness and health-seeking behavior). To account for prescribing norm of physicians, we use $p$ to denote the likelihood that each patient at presentation is prescribed antibiotic treatment, $0\le  p \le 1$. Thus, the overall prescribing rate, $p\theta$, mediates the selection pressure on resistance that is attributed to collective consequence of prescription behavior. 

We use the next-generation approach to calculate basic reproductive ratios for both strains, $\mathcal{R}_s$ and $\mathcal{R}_m$ in closed-form (see \emph{SI} for details). We assume that resistance is costly in the absence of treatment, but confers an advantage in the presence of treatment; that is, resistance compromises the efficacy of treatment, $0 \le \epsilon_m <1$. Comparing $\mathcal{R}_s$ and $\mathcal{R}_m$ allows us to answer questions of interest, such as predicting whether resistance can evolve in the long run.

For simplicity, we first consider resistance evolution under full treatment ($p = 1$), in which infected individuals, once seen by medical professionals at their presentation, unvaryingly receive antibiotic treatment. As shown in Fig. 2A, we characterize the conditions for resistance evolution in the parameter space $(\theta, \epsilon_m)$. For small $\theta$ values below a threshold (blue region in Fig. 2A), neither can the disease be eradicated, nor can resistance evolve. Disease can be eradicated for high $\theta$ and $\epsilon_m$ (yellow region in Fig. 2A). However, for combinations of intermediate $\theta$ and low $\epsilon_m$ (red region in Fig. 2A), resistance evolves and leads to disease escape despite full treatment.

To further gain intuitive understanding of how resistance evolution depends on antibiotic use, we plot the disease prevalence with respect to treatment probability $p$, corresponding to the three scenarios as colored in Fig. 2A. The sensitive strain is predominant for all $0\le p \le 1$, whereas the resistant strain is maintained at low frequency purely by the mutation-selection equilibrium (Fig. 2B). Disease can be eradicated for sufficiently high treatment rate and resistance has no chance to evolve (Fig. 2C). In Fig. 2D, disease eradication is impossible due to the emergence of resistance that greatly comprises the efficacy of treatment; resistance can be selected for $p$ above a critical threshold $p > p_h$ (see \emph{SI}), and as a consequence, the predominant incidence of infections switches from sensitive to resistant strains. 

Let us now turn our attention to this last scenario where resistance evolution is inevitable for $p > p_h$. Empirical evidence shows that there exists a threshold in prescription rate above which sustained resistance can cause huge public health crisis~\cite{Baquero_JAC96}. To determine the population optimum of antibiotic use, we need to take into account the impact of resistance on the cost-benefit analysis of antibiotic treatment (Fig. 3). The cost of sensitive infection, if treated, can be mitigated. In contrast, resistant infections may greatly exacerbate the overall cost for both treated and untreated cases~\cite{Roberts_CID09}. Under these conditions, the overall social burden of the disease can be minimized at $p = p_h$ (i.e., population optimum). Although the disease prevalence and thus the risk of infection for susceptible individuals can be lowered by overprescribing beyond $p_h$ (Fig. 3A), the cost associated with resistance is much greater than the benefit, if any, that full treatment could provide (Fig. 3B). 

Despite these population-level considerations, individual self-interest can cause antibiotic overuse, thereby leading to a tragedy of the commons. This is largely due to the disconnect between individual behavior and population-level resistance in prescribing decision-makings. Therefore, curbing antibiotic overuse requires provision of feedback to individual prescribing behavior with the social costs and consequences of their collective action. In light of this, we investigate whether population optimum of antibiotic use can be reached if the society learns from the collective consequences of treatment decisions and costs that are associated with resistance and accordingly adjusts prescription behavior. 

We assume disease dynamics coevolves with a social norm that governs prescription behavior (see Methods \& Model). We use evolutionary game theory to study the evolution of prescription norm~\cite{Nowak06,Sigmund_IR10}. Prescription norm changes in response to the actual payoffs of individual prescribing versus non-prescribing behavior, which are determined by disease prevalence and resistance evolution on the population level. This feedback loop between prescription behavior and resistance evolution constitutes an adaptive social learning process in which the society adjusts antibiotic use in response to the underlying resistance evolution. 

We find that how swiftly the society responds to the underlying resistance evolution has an impact on the coevolutionary dynamics (Fig. 4). Slow social learning leads to prolonged oscillatory dynamics of overprescribing and underprescribing, and thus gives chance for resistance to accumulate and build up in the population, causing resurgences of marked resistance prevalence alternated with sensitive infections (Fig. 4A). In stark contrast, fast social learning can help the population steer out cyclic oscillations of antibiotic use due to overcorrection. In this latter case, the society adapts prescription norm so quickly that resistance has no chance to grow into pronounced prevalence as it is outpaced by the change in prescribing behavior. Besides, fast social learning helps the society settle on a social norm that reaches the population optimum of antibiotic use (Fig. 4B).

We demonstrate that social learning without centralized institutions can maneuver the population towards a socially optimal policy of prescribing, therefore helping curb the overuse of antibiotics. Our theoretical results are in line with recent trial findings that highlight the importance of provision of social norm feedback in reducing antibiotic overuse~\cite{Hallsworth_Lancet16}. Taken together, in order to reach sustainable use of antibiotics, it is important to promote awareness of the population problem of resistance by providing prompt feedback to prescribing behavior with the social cost of resistance.

Owing to the drastic slowdown of new drug discoveries~\cite{Scannell_Nature12,Verhoef_CBDD15}, managing resistance evolution with an emphasis on human factors, as we demonstrate here, seems to be necessary and feasible~\cite{Levin_CID01}. Prior studies suggest that the consumption of antibiotics and the patterns in which different agents are deployed directly impact the frequency of resistance and the number of ineffective antibiotics~\cite{Baquero_JAC96,Seppala_NEJM97,Kristinsson_MDR97}. To inform rational use of antibiotics, efforts should be focused on developing new diagnostic technologies and strategies for reducing the inappropriate use of antibiotics~\cite{Kanthor_Nature14}, determining the optimal timing of deployment sequence for existing drugs~\cite{Wang_PNAS06,McClure_PRSB14}, and optimizing combination therapies~\cite{Bonhoeffer_PNAS97,Cottarel_TB07}. Moreover, promoting and enforcing infection control procedures in hospitals can prevent the spread of resistance and mitigate the impact of resistance on society~\cite{Nyquist_Jama98,Palumbi_Science01,Smith_PNAS05,Zur_LancetID11,Kanthor_Nature14,McClure_PRSB14,Rossolini_COP14,WHO15,Levin_CID01}. Along these lines, it is worthwhile for future study to incorporate population structure~\cite{Leventhal_NC15} and multiple drugs~\cite{Lipsitch_EID02,Donskey_NEJM00,Enne_Lancet01} in the coevolutionary dynamics of prescribing behavior and multi-drug resistance.

The socially optimum use of antibiotics implies a second order of dilemma -- not every sickness should be treated, but who on earth deserves the treatment and who would have to forgo?  This consideration leads to the ethics dilemma of accessibility of antibiotics, an important topic worthy of further investigation. Reducing antibiotics usage via national guidelines has been found to lead to significant decreases in resistance~\cite{Seppala_NEJM97,Kristinsson_MDR97}, yet denials or approvals of antibiotic treatment seem to be determined by an arbitrary trade-off between preventing resistance and treating infected patients~\cite{Projan_COM03}. With multiple interest groups such as pharmaceutical industry, public institutions as well as patients themselves involved in the problem, it is of fundamental interest to look into the issues of supervising the common pool resources and enhancing collaborative efforts through the behavioral perspective~\cite{WBG_book,Conly_BWHO10}.

\section{Methods \& Model}

The coevolutionary dynamics of prescription behavior and resistance is described by the following system of ordinary differential equations:
\begin{equation}\begin{cases}
	\frac{dS}{dt} =  b - \beta_sS[I^0_s + (1 - \epsilon_s)I^t_s] - \beta_mS[I^0_m + (1 - \epsilon_m)I^t_m] - dS,\\ 
	\frac{dI^0_s}{dt} = \beta_s(1 - \mu_s)S[I^0_s + (1 - \epsilon_s)I^t_s] + \beta_m\mu_mS[I^0_m + (1 - \epsilon_m)I^t_m] - p\theta I^0_s - \gamma^0_sI^0_s - dI^0_s,\\
	\frac{dI^t_s}{dt} = p\theta I^0_s - \gamma^t_sI^t_s - dI^t_s,\\
	\frac{dI^0_m}{dt} = \beta_s\mu_sS[I^0_s + (1 - \epsilon_s)I^t_m] + \beta_m(1-\mu_m)S[I^0_m+(1-	\epsilon_m)I^t_m] - p\theta I^0_m - \gamma^0_mI^0_m - dI^0_m,\\
	\frac{dI^t_m}{dt} = p\theta I^0_m - \gamma^t_mI^t_m - dI^t_m,\\
	\frac{dR_s}{dt} = \gamma^0_sI^0_s + \gamma^t_sI^t_s - dR_s,\\
	\frac{dR_m}{dt} = \gamma^0_mI^0_m + \gamma^t_mI^t_m - dR_m,\\
	\frac{dp}{dt} = \omega p(1 - p)(f_A - f_B).
\end{cases}\end{equation}
Here, $S$, $I$ and $R$ are the fractions of susceptible, infected and recovered individuals in the population with $I^0_s$, $I^t_s$, $I^0_m$ and $I^t_m$ denoting the four infection cases (infected with sensitive or resistant strains, and untreated or treated); $R_s$ and $R_m$ the two recovery cases, respectively. The two parameters $\mu_s$ and $\mu_m$ indicate the mutation rates between the two strains. The parameter $b$ denotes the birth rate per capita (which is set to be equal to the death rate $d$); $\beta_s$ and $\beta_m$ are the transmission rates of the two strains; $\gamma^0_s$, $\gamma^t_s$, $\gamma^0_m$ and $\gamma^t_m$ are the respective recovery rates of different infection cases (infected with sensitive or resistant strain; untreated or treated); $\epsilon_s$ and $\epsilon_m$ are the efficacies of the antibiotic for the two strains. 

We use the replicator equation to describe the evolution of prescribing behavior as given in the last equation~\cite{Nowak06}. We begin our analysis with characterizing the conditions for resistance evolution for given levels of antibiotic use $p$ in the long run. 

\subsection{Basic reproductive ratios}
The basic reproductive ratios $\mathcal{R}_s$ and $\mathcal{R}_m$ of the two strains can be determined by the spectral radius of the next-generation operator $FV^{-1}$, where $F$ is the reproduction matrix and $V$ the state transition matrix~\cite{Heffernan_JRSI05,Diekmann_JRSI09,Hurford_JRSI10}. We obtain that
\begin{eqnarray}
\mathcal{R}_s &=& \frac{\beta_s(\gamma^t_s + b) + \beta_s(1 - \epsilon_s)p\theta}{(\gamma^t_s + b)(p\theta + \gamma^0_s + b)},\label{BRRs}\\
\mathcal{R}_m &=& \frac{\beta_m(\gamma^t_m + b) + \beta_m(1 - \epsilon_m)p\theta}{(\gamma^t_m + b)(p\theta + \gamma^0_m + b)}.
\label{BRRm}
\end{eqnarray}

We assume that resistance incurs a fitness penalty, so that the transmission rates $\beta_s > \beta_m$. Despite the fact that laboratory studies revealed a scenario where compensatory evolution (resistant bacteria ameliorating the costs by acquiring fitness-compensatory mutations) and cost-free resistances can slow down the primary driver for reversibility and that co-selection between the resistance mechanism and other selected markers can delay any latent reversibility driven by fitness costs \textit{in vitro} ~\cite{Andersson_NR10}, clinical studies have found that the compensatory adaptation is not effective \textit{in vivo}, which is in line with our assumption~\cite{Maclean_EMPH15}. Theoretical arguments and experimental results, in addition, provide basis for that the fitness costs of resistance is critical to the displacement of resistant strains with sensitive ones~\cite{Andersson_NR10,Maclean_EMPH15}.

Both $\mathcal{R}_s(p)$ and $\mathcal{R}_m(p)$ are decreasing functions of $p$ under our model assumptions (see \textit{SI}). We demonstrate that the graphs of the two basic reproductive numbers will vary with values of their endpoints $\mathcal{R}_s(1)$ and $\mathcal{R}_m(1)$ (see the \textit{SI} for details). Without loss of generality, we scrutinize the following three cases: (\rom{1}) $R_s > 1$; (\rom{2}-a) $R_s(1) < 1$ and $R_m(1) < 1$; (\rom{2}-b) $R_s(1) < 1 < R_m(1)$. The case noteworthy in practice is the last one, in which resistant strains predominate and even full treatment can not eradicate the disease.

For a better understanding of the disease dynamics in case \rom{2}-b, we perform a further investigation of the relations among those parameters, where the presentation rates $\theta_s$ and $\theta_m$ for the two infected cases are seen as independent. Substituting $p = 1$ into \eqref{BRRs} and $\eqref{BRRm}$ and combining with the inequality $R_s(1) < 1 < R_m(1)$, we derive an equivalent condition for case \rom{2}-b to occur:
\begin{equation}
	\begin{cases}
	0 \leq \epsilon_s \leq 1, \quad \text{if} \> \theta_s > \theta^*_s,\\
	\epsilon^*_s < \epsilon_s \leq 1, \quad \text{if} \> \theta_s < \theta^*_s.
	\end{cases}
	\text{and} \>
	\begin{cases}
	0 \leq \epsilon_m \leq 1, \quad \text{if} \> \theta_m < \theta^*_r,\\
	0 \leq \epsilon_m < \epsilon^*_r, \quad \text{if} \> \theta_m >  \theta^*_r.
	\end{cases}
	\end{equation}
	The values of $\theta^*_s$, $\epsilon^*_s$, $\theta^*_r$ and $\epsilon^*_r$ are given in the \textit{SI}. Therefore, to get the basic reproductive ratio below $1$ and thus control the disease, we need $\theta_m$ to be greater than $\theta_m^\ast$ or $\epsilon_m$ greater than $\epsilon_m^\ast$. The two solvents correspond to either patients presenting promptly after infection or introducing potent antibiotics in treatment.

Moreover, let $p_h$ be the critical prescribing probability (which can be translated into treatment coverage, namely, the proportion of patients that are prescribed antibiotic treatment) at which the dominance of the two strains switches in case \rom{2}-b. For $\gamma_s^0 = \gamma_m^0$, we derive a simplified form 
\begin{equation}
	p_h = \frac{\beta_s - \beta_m}{\theta[\frac{\beta_m(1 - \epsilon_m)}{\gamma^t_m + b} - \frac{\beta_s(1 - \epsilon_s)}{\gamma^t_s + b}]},
	\end{equation}
where $p_h$ is referred to as the social optimum of antibiotic use (see details in the \emph{SI}).

\subsection{Cost-benefit analysis}
 When $p < p_h$, the sensitive strain dominates and the system converges to an equilibrium, of which we derive a closed form approximation $(\hat{S}, \hat{I^0}_s, \hat{I^t}_s, \hat{R}_s)$; when $p > p_h$, the resistant strain dominates, with the system converging to another equilibrium, approximated by $(\hat{S}, \hat{I^0_m}, \hat{I^t_m}, \hat{R_m})$. We prove in the \textit{SI} that $S$ is increasing while $I^0$ and $I^0 + I^t$ are decreasing with respect to $p$ for both equilibria. 
 
For susceptible individuals, if the sensitive strain dominates, the infection probability at equilibrium is approximately
	\begin{equation}
	\varphi_s = \frac{\beta_s[\hat{I}^0_s + (1 - \epsilon_s)\hat{I}^t_s]}{\beta_s[\hat{I}^0_s + (1 - \epsilon_s)\hat{I}^t_s]+b},
	\end{equation}
	which can be simplified as $1 - \hat{S} = 1 - \frac{1}{\mathcal{R}_s}$.  Analogously, if the resistant strain prevails, the infection probability at equilibrium is approximately
	\begin{equation}
	\varphi_m = \frac{\beta_m[\hat{I}^0_m + (1 - \epsilon_m)\hat{I}^t_m]}{\beta_m[\hat{I}^0_m + (1 - \epsilon_m)\hat{I}^t_m]+b} = 1 - \frac{1}{\mathcal{R}_m}.
	\end{equation}
Although the actual treatment cost can be determined only after treatment outcomes, it is expected that (1) sensitive strain claims lower sickness and treatment costs while it redounds to greater treatment benefit ($C_{I_s} < C_{I_m}$, $B_{T_s} > B_{T_m}$ and $C_{T_s} < C_{T_m}$), (2) treatment of patients with sensitive strains can mitigate the overall cost of infection ($C_{T_s} - C_{I_s}< B_{T_s}$), and in contrast (3) treatment of patients with resistant strains may exacerbate the overall cost of infection ($C_{T_m} - C_{I_m} > B_{T_m}$)~\cite{Roberts_CID09}.
	
When it comes to the population, the total social cost is a piecewise function
\begin{equation}
	C_{\text{social}} = 
	\begin{cases}
	(C_{T_s} - B_{T_s})(\hat{I}^0_s + \hat{I}^t_s) + (C_{I_s} + B_{T_s} - C_{T_s})\hat{I}^0_s, \qquad \text{when} \> p < p_h\\
	(C_{T_m} - B_{T_m})(\hat{I}^0_m + \hat{I}^t_m) - (C_{T_m} - C_{I_m} - B_{T_m})\hat{I}^t_m. \qquad \text{when} \> p > p_h.
	\end{cases}
	\end{equation}
Invoking the monotonicity of $\hat{I}^0_s$ and $\hat{I}^0_s + \hat{I}^t_s$, it is easy to verify that the total social cost $C_{\text{social}}$ is decreasing when $p < p_h$. However, $C_{\text{social}}$ may not be monotonic when $p > p_h$.

\subsection{Social learning}
We consider that prescription behavior coevolves with disease dynamics. We regard the problem as a two strategy game, prescribing (denoted as $A$) vs nonprescribing ($B$). The evolution of prescribing behavior can be described by:
	\begin{equation}
	\dot{p} = \omega p(1-p)(f_A - f_B),
	\end{equation}
where $p$ is the frequency of prescribing $A$ and $\dot{p}$ is referred to as the rate of prescription norm evolution, driven by the time scale parameter of social learning, $\omega$.	

The expected payoffs $f_A$ and $f_B$ are
	\begin{eqnarray}
	f_A &=& \lambda_s(B_{T_s} - C_{T_s}) + 
	\lambda_m(B_{T_m} - C_{T_m}),\\
	f_B &=& \lambda_s(- C_{I_s}) + 
	\lambda_m(- C_{I_m}),
	\end{eqnarray}
	with
	\begin{eqnarray}
	\lambda_s &=& \frac{\beta_s[I^0_s + (1 - \epsilon_s)I^t_s]}{\beta_s[I^0_s + (1 - \epsilon_s)I^t_s] + \beta_m[I^0_m + (1 - \epsilon_m)I^t_m]},\\
	\lambda_m &=& \frac{\beta_m[I^0_m + (1 - \epsilon_m)I^t_m]}{\beta_s[I^0_s + (1 - \epsilon_s)I^t_s] + \beta_m[I^0_m + (1 - \epsilon_m)I^t_m]}.
	\end{eqnarray}
Here $\lambda_s$ and $\lambda_m$ are the conditional probabilities of individuals being infected with sensitive or resistant strains, respectively.
	
Denote the fractions of individuals infected with sensitive and resistant strains by $I_s$ and $I_m$, respectively ($I_s = I^0_s + I^t_s$ and $I_m = I^0_m + I^t_m$). The behavior of the disease dynamics are described by $I_s$ and $I_m$ while the prescribing norm is presented by $p$, all as functions of the time $t$. 

\section*{Acknowledgements}
We are grateful for support from the G. Norman Albree Trust Fund, Dartmouth Faculty Startup Fund, Walter \& Constance Burke Research Initiation Award and NIH Roybal Center Pilot Grant.

\clearpage

\begin{figure}[htbp]
   \centering
   \includegraphics[width=\columnwidth]{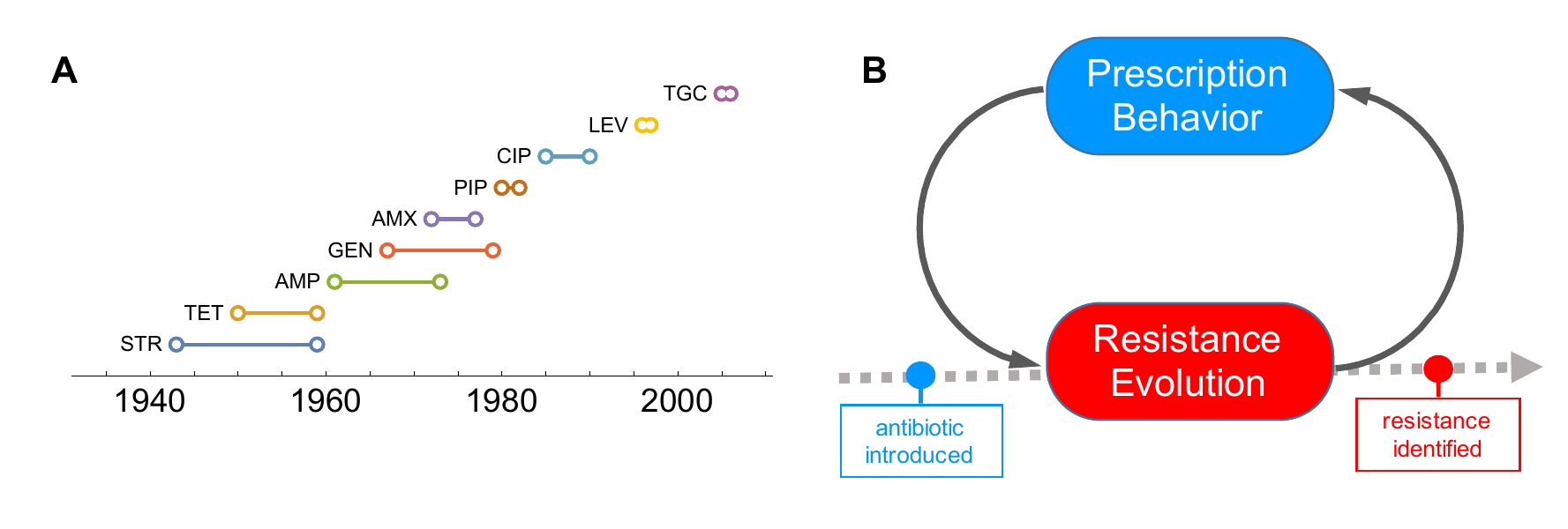} 
      \caption{Problem of antibiotic resistance. (A) Timeline plot of resistance emergence for common antibiotics. The time period of an antibiotic's effectiveness from its introduction to first resistance identified becomes increasingly short. (B) Prescribing behavior is one of the driving factors contributing to the fast emergence of resistance. The interaction between resistance evolution and prescripting behavior plays an important role in determining the timeline of resistance emergence.}
   \label{fig1}
\end{figure}
\newpage

\begin{figure}[htbp]
   \centering
   \includegraphics[width=\columnwidth]{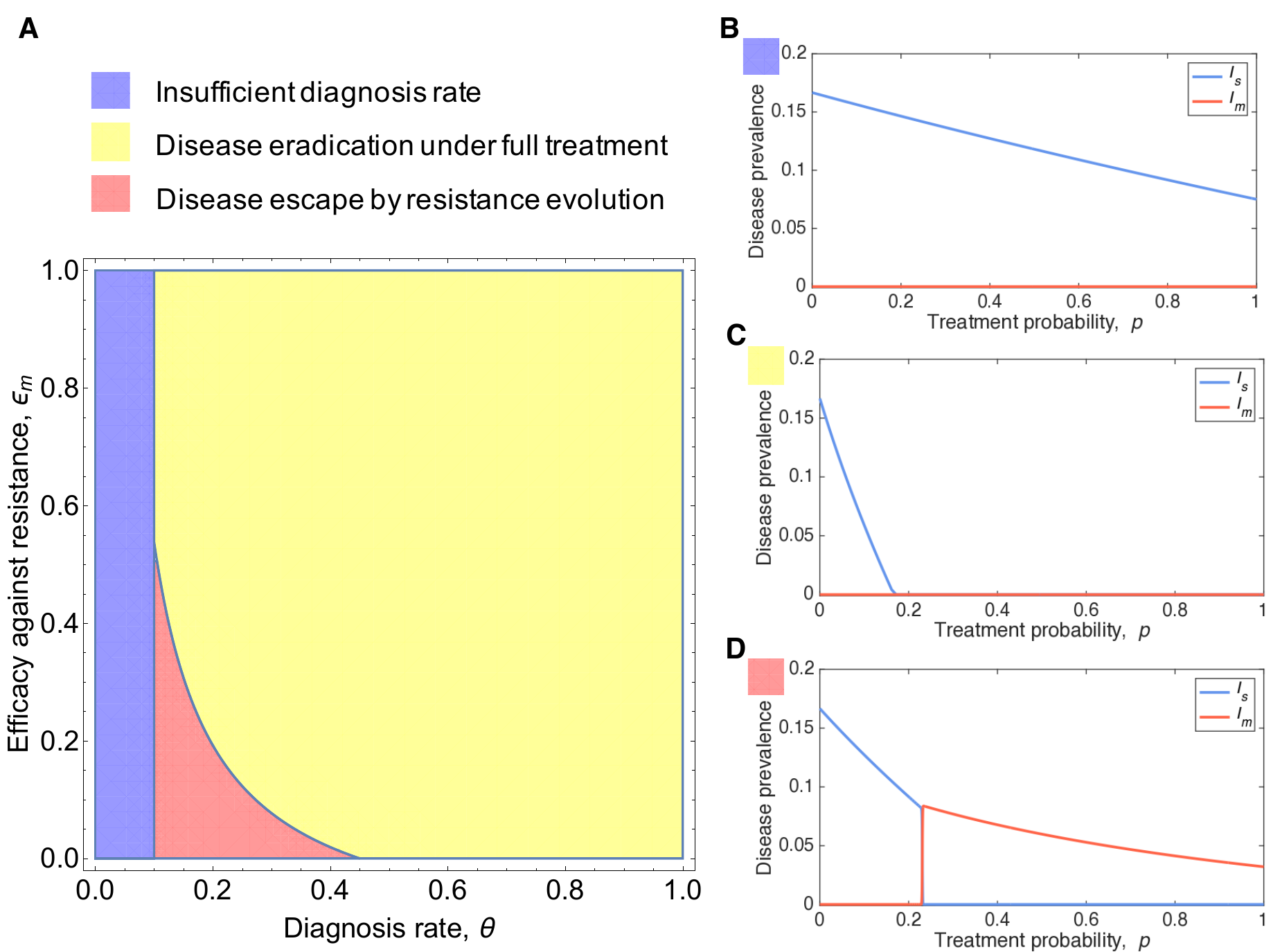} 
      \caption{Resistance evolution and antibiotic usage. (A) Shown is the parameter region of presentation rate $\theta$ and efficacy against resistance $\epsilon_m$ under which antibiotic resistance can emerge. (B), (C), (D) plot the prevalence of sensitive and resistant strains as a function of treatment probability $p$. Resistance can be favored under sufficiently high treatment coverage and as a consequence, the tragedy of the commons in antibiotic overuse can occur. Parameters:
      $b = 0.1$,
      $\beta_s = 0.3$,
      $\beta_m = 0.26$,
      $\gamma_s^0 = \gamma_m^0 = 0.1$,
      $\gamma_s^t = 0.3$,
      $\gamma_m^t = 0.2$,
      $\epsilon_s=1$,
      (A) $\epsilon_m = 0$,
      (B) $\epsilon_m = 0.6$, $\theta = 0.05$,
      (C) $\epsilon_m = 0.8$, $\theta = 0.6$,
      (D) $\epsilon_m = 0$, $\theta = 0.2$,
      (B)-(D) $\mu_s = \mu_m = 10^{-6}$.
      }
   \label{fig2}
\end{figure}

\newpage

\begin{figure}[htbp]
   \centering
   \includegraphics[width=.9\columnwidth]{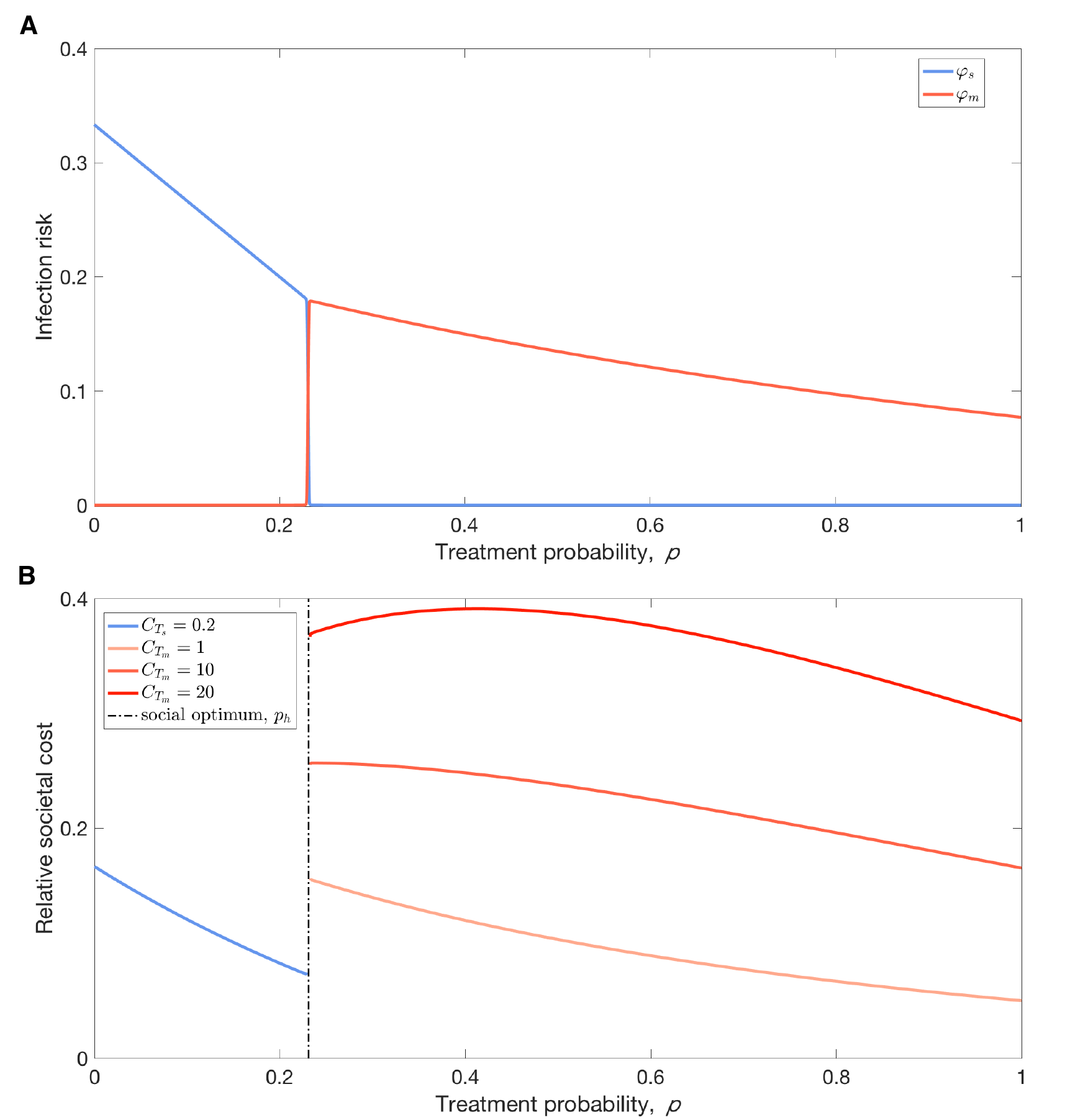} 
      \caption{Cost-benefit analysis of antibiotic usage. (A) From the perspective of susceptible individuals, their infection risk reduces with increasing treatment probability (coverage) $p$, but there exists a threshold of treatment coverage, $p_h$, above which the risk of infection almost exclusively comes from resistant strain instead of sensitive stain. (B) Accounting for the differing costs of treating patients infected with sensitive and resistant strains, the threshold $p_h$ corresponds to the social optimum of antibiotic usage: the total societal disease burden is decreasing with $p$ for $p < p_h$, but followed by a surge in the total cost of infection due to resistance. 
    Parameters:
    (A),(B) $b = 0.1$,
      $\beta_s = 0.3$,
      $\beta_m = 0.26$,
      $\gamma_s^0 = \gamma_m^0 = 0.1$,
      $\gamma_s^t = 0.3$,
      $\gamma_m^t = 0.2$,
      $\epsilon_s=1$, $\epsilon_m = 0$,
      $\theta = 0.2$,
      $\mu_s = \mu_m = 10^{-6}$,
      (B) relative social burden of the disease: $C_{I_s} = 1 $, $C_{T_s} = 0.2$, $B_{T_s} = 0.3$,
         $C_{I_m} = 2$, $B_{T_m} = 0.1$.
      }
   \label{fig3}
\end{figure}
\newpage

\begin{figure}[htbp]
   \centering
   \includegraphics[width=.9\columnwidth]{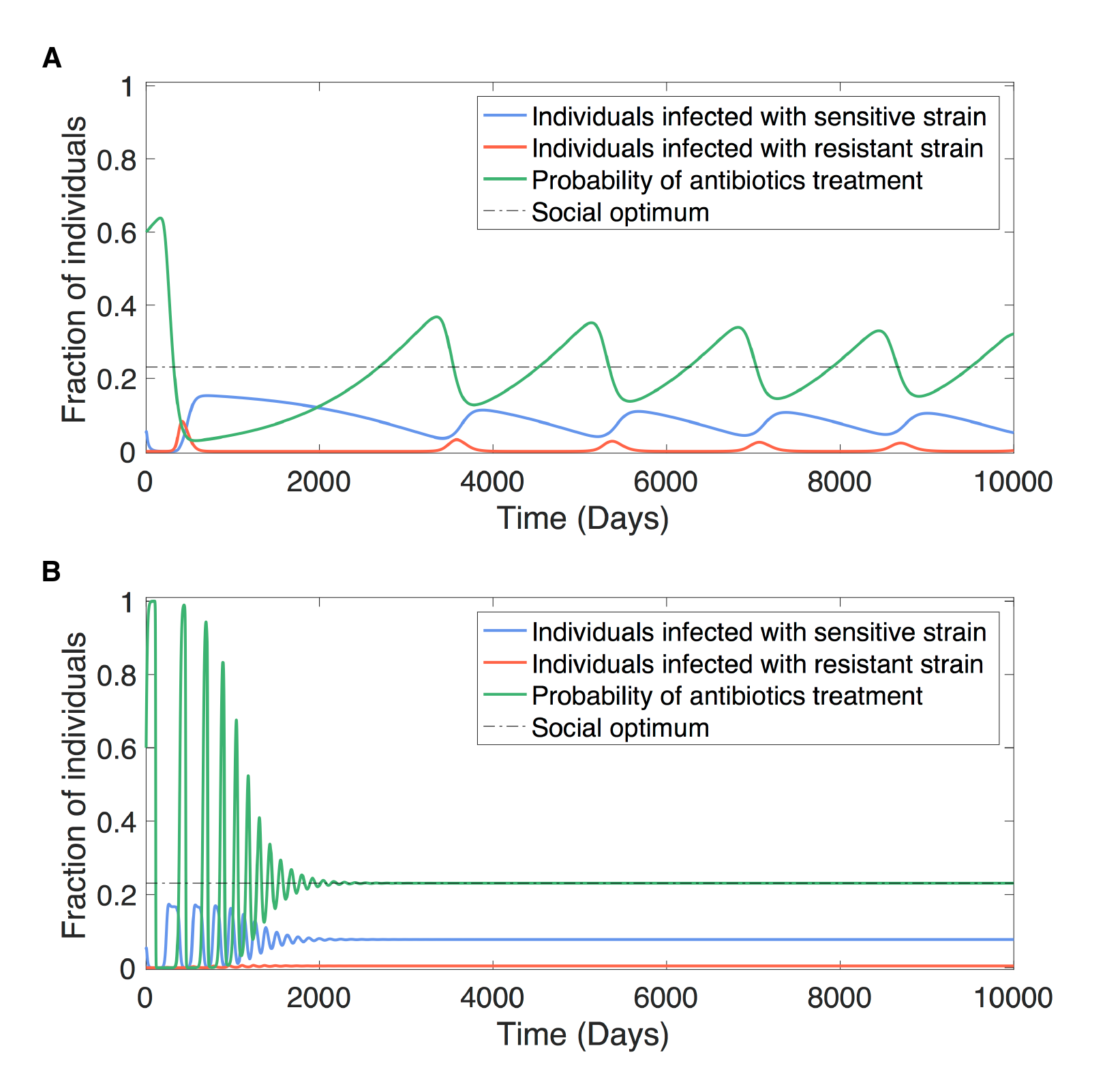} 
      \caption{Social learning impacts antibiotic usage. The antibiotic usage behavior can be adjusted in response to the underlying dynamics of resistance evolution. The convergence of optimum antibiotic usage is determined by how promptly the society learns from the collective consequences of treatment decisions and costs that are associated with resistance. (A) Slow learning leads to oscillatory dynamics of antibiotic usage (between overuse and lesser use) together with alternating dominance of resistant and sensitive strains. (B) Fast learning can steer out cyclic oscillations of antibiotic usage due to overcorrection, and therefore helps the society quickly reach socially optimal usage. Parameters:  
      $b = 0.1$,
      $\beta_s = 0.3$,
      $\beta_m = 0.26$,
      $\gamma_s^0 = \gamma_m^0 = 0.1$,
      $\gamma_s^t = 0.3$,
      $\gamma_m^t = 0.2$,
      $\epsilon_s=1$, $\epsilon_m = 0$,
      $\theta = 0.2$,
      $\mu_s = \mu_m = 10^{-6}$,
     relative social burden of the disease: $C_{I_s} = 1 $, $C_{T_s} = 0.2$, $B_{T_s} = 0.3$,
         $C_{I_m} = 2$, $C_{T_m}=20$, $B_{T_m} = 0.1$, (A) $\omega = 0.001$, (B) $\omega = 0.1$.
      }
   \label{fig4}
\end{figure}

\clearpage

\end{document}